\documentclass{article}
\usepackage{emulateapj}
\usepackage{onecolfloat}
\usepackage{graphicx} 
\submitted{To be published in ApJ}
\newcommand\gtsima{$\scriptscriptstyle \; \buildrel > \over \sim \;$}
\newcommand\etal{{et al.}}

\newcommand\teff{T$_{\rm eff}$}
\newcommand\logg{log $g$}
\newcommand\rss{$r_{ss}$}
\begin{document}
\twocolumn [
\title{The r-process in the early Galaxy}

\author{Jennifer A. Johnson}

\affil{OCIW, 813 Santa Barbara St., Pasadena, CA 91101}
\authoremail{jennifer@ociw.edu}

\and

\author{Michael Bolte}
\affil{UCO/Lick Observatory, University of 
 California, Santa Cruz,
CA~95064} 
\authoremail{bolte@ucolick.org}

\begin{abstract}

We report Sr, Pd and Ag abundances for a sample of metal-poor field
giants and analyze a larger sample of Y, Zr, and Ba abundances.  The
$\lbrack$Y/Zr$\rbrack$ and $\lbrack$Pd/Ag$\rbrack$ abundance ratios
are similar to those measured for the r-process-rich stars CS
22892-052 and CS 31082-001. The $\lbrack$Pd/Ag$\rbrack$ 
ratio is larger than predicted from the
solar-system r-process abundances.  The constant
$\lbrack$Y/Zr$\rbrack$ and $\lbrack$Sr/Y$\rbrack$ values in the field
stars places strong limits on the contributions of the weak s-process
and the main s-process to the light neutron-capture elements.  Stars
in the globular cluster M 15 possess lower $\lbrack$Y/Zr$\rbrack$
values than the field stars.  There is a large dispersion in
$\lbrack$Y/Ba$\rbrack$.  Because the r-process is responsible for the
production of the heavy elements in the early Galaxy, these
dispersions require varying light-to-heavy ratios in
r-process yields.

\end{abstract}

]

\section{Introduction}
Burbidge \etal{} (1957) and Cameron (1957) showed that only two sets
of physical conditions were necessary to explain the abundances of the
heavy (A $>$ 65) elements in the solar system. Because of strong
Coulomb forces, the build-up of heavier nuclei happens through
neutron-capture. The first of these neutron-capture processes is the
s-process, where neutron captures onto seed nuclei take place much
more slowly than $\beta -$decays.  The s-process is thought to take
place in two distinct environments.  The `main' s-process occurs in
low-mass AGB stars, while the `weak' s-process occurs during helium
burning in massive stars. While the main s-process can make
neutron-rich material up to $^{209}$Bi (Clayton \& Rassbach 1967), the
weak s-process is not predicted to make significant amounts of
material with A$> 90$ (Couch, Schmiedekamp, \& Arnett 1974).  The
second neutron-capture process, the r-process, takes place when
conditions are such that neutron capture rates are much higher than
$\beta-$decay rates. It produces a distinctive pattern in the
abundance ratios, the most noticeable features being the so-called
r-process peaks. These peaks, at A$\sim$80, 130, and 196, are the
signatures of nucleosynthesis events which reached the neutron magic
numbers of 50, 82, and 126.

Despite having been identified as taking place in an environment with
 rapid neutron captures, the astrophysical phenomena that create the
 r-process remain unidentified.  The neutrino wind in Type II SN
 showed promise (Woosley \& Hoffman 1992; Woosley \etal{} 1994), but
 two problems arose. First, there remain questions about whether the
 entropy in the wind is sufficiently high to produce the r-process
 (Takahashi, Witti, \& Janka 1994; Qian \& Woosley 1996, but see
 Otsuki \etal{} 2000; Wanajo \etal{} 2001).  Also, Freiburghaus
 \etal{} (1999a) found that high-entropy wind models cannot produce an
 r-process pattern for A$<$110 because those elements are synthesized
 during the low-entropy, neutron-deficient $\alpha$-rich freezeout
 phase of the wind. Witti, Janka, \& Takahashi (1994) and Woosley
 \etal{} (1994) also show that the elements near N=50 are overproduced
 relative to the more massive nuclei in neutrino wind models. So
 either the neutrino-wind is not the source of any r-process products,
 or the conditions are such that the material with A$<$110 is either
 not ejected or is diluted. Models have shown that merging neutron
 stars may be the source of significant amounts of r-process material
 (e.g. Lattimer \& Schramm 1974; Rosswog \etal{} 2000).  Freiburghaus,
 Rosswog \& Thielemann (1999b) did parameterized calculations of the
 nucleosynthesis in the ejecta and found that it could be a source for
 nuclei with A\gtsima130.  An earlier suggestion that the r-process
 occurred in helium-burning regions, either cores of low-mass stars or
 in the helium shell in SN, was rejected because unacceptably large
 amounts of $^{13}$C were required to make the A$\sim$195 peak (Cowan,
 Cameron \& Truran 1985).  However, helium-burning phases can still
 make r-process isotopes in the range A$\sim$80 with about half as
 much $^{13}$C. Truran, Cowan, \& Fields (2001) updated the
 calculations for the helium-shell shock r-process and find that for
 normal amounts of C, it can provide interesting amounts of material
 with A $<$ 130.

Abundances in metal-poor stars provide insight into the r-process
because they bear the marks of relatively few nucleosynthesis events
and because the r-process is thought to be the sole source of many of
the heavy elements in the early Galaxy (Truran 1981). However,
alternative sources, such as the weak s-process, may contribute
substantial amounts to the lightest of the neutron-capture elements,
such as Sr, Y and Zr (e.g. Prantzos, Hashimoto \& Nomoto 1990).  In
recent years, much new information has become available on the
abundances of heavy elements in metal-poor stars (e.g. Gilroy \etal{}
1988; McWilliam \etal{} 1995; Ryan \etal{} 1996; Westin \etal{} 2000;
Burris \etal{} 2000), revealing a wide diversity in some abundance
ratios, such as [Sr/Ba] and a remarkable consistency in others, such
as [Ba/Eu]. The information on the intermediate-mass elements such as
Pd and Ag is particularly interesting, since the abundance ratios in
one star, CS 22892-052, showed a larger difference between the
abundances of the odd-Z and even-Z elements than seen in the solar
system (Sneden \etal{} 2000a).

Johnson (2002) (Paper I) presented the abundances of up to 17 neutron-capture
elements in a sample of 22 metal-poor ([Fe/H] $< -1.7$) field red giants.
In this paper we add Sr abundances for these stars, as well as Pd and
Ag abundances for three stars from this sample.
We then analyze the abundance patterns for all the heavy elements
in the Paper I sample to learn more about 
the r-process in the early Galaxy.

\section{Observations and Data Reduction}

The full details of the observations and data reduction are in Paper I.
Briefly, we obtained high-resolution spectra on two echelle spectrographs.
We observed 12 stars with HIRES (Vogt \etal{} 1994) 
on Keck I in May and June 1997. These data cover from 3200\AA{} to
4700\AA{} with R$\sim$45,000. The S/N was $\sim$ 200 at 4000\AA.
We also observed 21 stars, including 11 of the 12 HIRES stars, with 
the Hamilton on the Shane 3-meter at Lick Observatory (Vogt 1987).
The Hamilton data have larger
wavelength coverage (3800\AA{} to 7000\AA{}) and higher resolution
(R$\sim$60,000), but lower S/N ($\sim$100 at 6000\AA{}). 
We measured over 7000 
equivalent widths (EWs) and synthesized over 200 additional lines. 
Paper I reports the
abundances of 30 elements for our sample of stars, including 17
neutron-capture elements. 

We used Kurucz model
atmospheres\footnote{http://cfaku5.harvard.edu/}. The effective
temperature was chosen so that the abundance derived from \ion{Fe}{1} lines
did not depend on their excitation potential. Gravity was set from the
ionization balance from the \ion{Fe}{1}/\ion{Fe}{2} lines. The microturbulent
velocity ($\xi$) was changed until there was no slope in the
abundances vs. EW plot for the \ion{Fe}{1} lines.  The errors
shown in this paper caused by uncertainties in the atmospheric
parameters take into consideration random errors of $\pm$ 100K in
\teff, $\pm$ 0.3 dex in \logg, and $\pm$ 0.3 km/s in $\xi$ as well as
the scatter caused by inaccurate gf-values and EWs.  Errors for
abundance ratios consider the similarity of the two elements' response
to atmospheric model parameter changes in the manner discussed in
McWilliam \etal{} (1995) (Paper I). We note that when solar values are
used, we used the meteoritic values from Anders \& Grevesse (1989). 

However, three interesting elements were not considered in that
study, Pd, Ag and Sr. Below we derive abundances for these three
elements to bring to 20 the number of neutron-capture elements
with measurements in our metal-poor star sample.

\begin{figure}[b]
\begin{center}
\includegraphics[width=3in,angle=0]{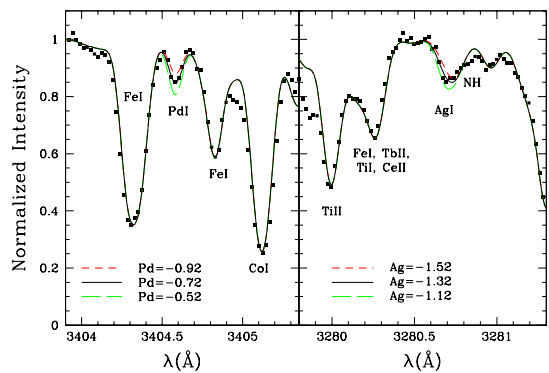}
\caption{\footnotesize Synthesis of the 3404\AA{} \ion{Pd}{1} line and
the 3280\AA{} \ion{Ag}{1} line in HD 186478 ([Fe/H] $=-2.60$. The black
solid line shows the best synthesis, while the other lines show
changes of $\pm$0.2 dex. For the Pd synthesis, the Fe abundance
has been increased 0.2 dex, the Co abundance decreased by $-$0.6 dex
and the Zr abundance increased by 0.3 dex over the abundances
derived in Paper I. For the Ag synthesis, the Zr abundance ($-$0.2 dex)
has been decreased. We also adopted [O/Fe] =1.00 dex. These changes are
not unexpected given the uncertainities in abundances and gf-values, but
they had no effect on the derived Pd and Ag abundances.} 
\end{center}
\end{figure}

\begin{figure}[t]
\begin{center}
\includegraphics[width=3in,angle=0]{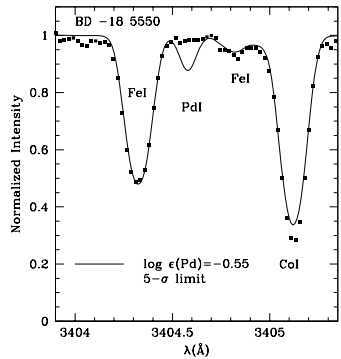}
\caption{\footnotesize Synthesis of the 3404 \AA{} \ion{Pd}{1} line
region in BD -18 5550 ([Fe/H]$=-3.04$). The solid black line
shows the 5-$\sigma$ upper limit on the Pd abundance.} 
\end{center}
\end{figure}

\begin{deluxetable}{rcrrrcrr}
\tablenum{1}
\tablewidth{0pt}
\tablecaption{Linelist near \ion{Ag}{1} at 3280 \AA}
\tablehead{
\colhead{$\lambda$} & \colhead{Species} & \colhead{E.P.}& 
\colhead{log $gf$} & \colhead{$\lambda$} & \colhead{Species} & \colhead{E.P.}& 
\colhead{log $gf$} \\
\colhead{\AA} & \colhead{} & \colhead{eV} & \colhead{} & 
\colhead{\AA} & \colhead{} & \colhead{eV} & \colhead{} 
}
\startdata
3279.010 &  CoI &   3.13 & $-$2.40 &3280.957 &  NH &   1.61 & $-$1.61 \\
3279.130 &  OH &   1.50 & $-$1.51 &3280.957 &  NH &   1.71 & $-$1.08 \\
3279.251 &  CoI &   1.96 & $-$0.85 &3280.957 &  NH &   1.71 & $-$1.10 \\
3279.255 &  OH &   1.64 & $-$1.85 &3281.001 &  NH &   1.82 & $-$1.05 \\
3279.266 &  ZrII &   0.09 & $-$0.23 &3281.044 &  NH &   2.16 & $-$1.63 \\
3279.325 &  ErII &   0.64 & $-$0.13 &3281.101 &   CeII &   0.88 & $-$0.58 \\
3279.733 &   FeI &   2.99 & $-$1.83 &3281.125 &   VII &   2.56 & $-$0.63 \\
3279.812 &  CuI &   1.64 & $-$2.17 &3281.141 &  NH &   1.82 & $-$0.57 \\
3279.839 &  CeII &   0.29 & $-$0.50 &3281.196 &  NH &   1.51 & $-$0.59 \\
3279.848 &  VII &   2.37 &  0.01 &3281.212 &  NH &   1.82 & $-$0.58 \\
3279.972 &  HfII &   0.45 & $-$1.14 &3281.304 &   FeII &   1.04 & $-$2.99 \\
3279.995 &  TiII &   1.12 & $-$0.84 &3281.328 &  NH &   1.51 & $-$0.61 \\
3280.003 &  UII &   0.11 & $-$1.19 &3281.328 &  NH &   1.51 & $-$0.62 \\
3280.093 &  DyII &   0.10 & $-$0.53 &3281.483 &   NdII &   1.60 &  0.07 \\
3280.155 &  OH &   1.64 & $-$2.25 &3281.600 &   CoI &   0.17 & $-$3.76 \\
3280.217 &  ErII &   0.05 & $-$0.67 &3282.247 &   CoI &   1.78 & $-$2.84 \\
3280.267 &   FeI &   3.30 & $-$0.22 &3282.310 &   GdII &   1.34 &  0.26 \\
3280.312 &  TbII &   0.00 &  0.13 &3282.334 &   TiII &   1.22 & $-$0.29 \\
3280.368 &  TiI &   1.07 & $-$2.25 &3282.447 &   FeI &   2.50 & $-$2.00 \\
3280.481 &  CeII &   0.55 & $-$0.07 &3282.480 &   UII &   0.00 & $-$1.31 \\
3280.536 &   RhI &   0.19 & $-$0.52 &3282.540 &   VII &   2.37 &  0.02 \\
3280.577 &   RhI &   0.43 & $-$0.87 &3282.689 &   NiI &   0.17 & $-$2.17 \\
3280.677 &   AgI &   0.00 & $-$0.48 &3282.705 &  OH &   1.98 & $-$1.39 \\
3280.678 &   AgI &   0.00 & $-$0.46 &3282.717 &   FeI &   2.95 & $-$2.08 \\
3280.683 &   AgI &   0.00 & $-$0.94 &3282.730 &   ZrI &   0.15 & $-$0.53 \\
3280.685 &   AgI &   0.00 & $-$0.96 &3282.766 &   NdII &   0.00 & $-$1.36 \\
3280.735 &   ZrII &   0.71 & $-$1.10 &3282.771 &  NH &   1.32 & $-$0.62 \\
3280.759 &  NH &   1.71 & $-$0.57 &3282.837 &   ZrII &   1.83 &  0.30 \\
3280.775 &   MnI &   2.14 & $-$2.23 &3282.858 &  NH &   1.32 & $-$0.64 \\
3280.775 &   FeI &   3.02 & $-$2.53 &3282.858 &  NH &   1.32 & $-$0.66 \\
3280.844 &   SmII &   0.10 & $-$1.09 &3282.892 &   DyII &   0.59 & $-$1.00 \\
3280.957 &  NH &   1.61 & $-$1.58 &3282.904 &   FeI &   3.27 & $-$0.61 \\
3280.957 &  NH &   1.61 & $-$1.60 &   &     &   &   \\

\enddata
\end{deluxetable}

\begin{deluxetable}{rcrrrcrr}
\tablenum{2}
\tablewidth{0pt}
\tablecaption{Linelist near \ion{Pd}{1} at 3404 \AA}
\tablehead{
\colhead{$\lambda$} & \colhead{Species} & \colhead{E.P.}& 
\colhead{log $gf$} & \colhead{$\lambda$} & \colhead{Species} & \colhead{E.P.}& 
\colhead{log $gf$} \\
\colhead{\AA} & \colhead{} & \colhead{eV} & \colhead{} & 
\colhead{\AA} & \colhead{} & \colhead{eV} & \colhead{} 
}
\startdata
3402.396 & CrI &   3.10 & $-$0.56 & 3405.096 & CoI &   0.43 & $-$1.56 \\
3402.424 & TiII &   1.22 & $-$1.00 & 3405.104 & CoI &   0.43 & $-$0.88 \\
3402.429 & FeI &   2.73 & $-$1.84 & 3405.112 & CoI &   0.43 & $-$1.48 \\
3402.462 & SmII &   0.38 & $-$0.56 & 3405.113 & CoI &   0.43 & $-$1.77 \\
3402.508 & OsI &   0.00 & $-$2.25 & 3405.119 & CoI &   0.43 & $-$0.99 \\
3402.634 & FeI &   2.76 & $-$2.14 & 3405.125 & CoI &   0.43 & $-$1.47 \\
3402.900 & ZrII &   1.53 & $-$0.33 & 3405.127 & CoI &   0.43 & $-$1.56 \\
3403.084 & SmII &   0.49 & $-$0.91 & 3405.131 & CoI &   0.43 & $-$1.15 \\
3403.271 & FeI &   2.83 & $-$1.93 & 3405.136 & CoI &   0.43 & $-$1.51 \\
3403.344 & TiI &   1.07 & $-$0.51 & 3405.138 & CoI &   0.43 & $-$1.48 \\
3403.345 & CrII &   2.43 & $-$0.54 & 3405.141 & CoI &   0.43 & $-$1.32 \\
3403.346 & MoI &   1.42 & $-$1.61 & 3405.145 & CoI &   0.43 & $-$1.62 \\
3403.597 & CeII &   0.52 & $-$0.79 & 3405.147 & CoI &   0.43 & $-$1.47 \\
3403.693 & ZrII &   1.00 & $-$0.60 & 3405.149 & CoI &   0.43 & $-$1.49 \\
3404.290 & FeI &   1.01 & $-$2.58 & 3405.152 & CoI &   0.43 & $-$1.84 \\
3404.290 & FeI &   2.73 & $-$0.67 & 3405.154 & CoI &   0.43 & $-$1.51 \\
3404.360 & FeI &   2.20 & $-$0.88 & 3405.155 & CoI &   0.43 & $-$1.63 \\
3404.580 & PdI &   0.81 &  0.32 & 3405.159 & CoI &   0.43 & $-$1.62 \\
3404.764 & FeI &   2.73 & $-$2.20 & 3405.159 & CoI &   0.43 & $-$1.64 \\
3404.830 & ZrII &   0.36 & $-$0.70 & 3405.161 & CoI &   0.43 & $-$1.84 \\
3404.902 & FeI &   2.69 & $-$2.60 & 3405.654 & DyII &   0.59 & $-$0.66 \\
3404.908 & CeII &   0.23 & $-$0.65 & 3405.838 & FeI &   2.69 & $-$1.86 \\
3404.961 & TiII &   1.22 & $-$2.72 & 3405.978 & CeII &   0.55 &  0.04 \\
3405.064 & TiI &   1.05 & $-$0.80 & 3406.437 & FeI &   3.27 & $-$0.88 \\
3405.079 & CoI &   0.43 & $-$2.77 & 3406.552 & FeI &   2.45 & $-$1.90 \\
3405.087 & CoI &   0.43 & $-$0.71 & 3406.800 & FeI &   2.22 & $-$1.10 \\
\enddata
\end{deluxetable}

\section{Abundances}
\subsection{Pd and Ag}

We could only use one Pd and one Ag line to measure abundances. Both
lines lie to the blue of 3500 \AA{} where the spectrum, even in
metal-poor stars, is very crowded. Therefore, we synthesized the
spectrum in these regions. Examples for the Pd and Ag lines are shown
in Figure 1. The initial line list was taken from Kurucz CD-ROM 23
(Kurucz \& Bell 1995) and modified when more recent gf-values could be
found. We also eliminated lines from the list that had no noticeable
contribution to the spectral synthesis. For the \ion{Pd}{1} line
list, we used the gf-values of O'Brian \etal{} (1991) for the
\ion{Fe}{1} lines at 3404.290\AA{} and 3404.360\AA{}. The \ion{Co}{1}
line at 3405.13\AA{} had hyperfine splitting (HFS) constants from
Pickering (1996). The Pd line has a gf-value from Bi\'emont \etal{}
(1982). This region of the spectrum is also littered with NH lines.
The gf-values and wavelengths for the NH lines were taken from
Kurucz\footnote{http://cfaku5.harvard.edu} and most have theoretically
predicted wavelengths and gf-values. Luckily, none of the lines near
Pd 3404\AA{} were strong enough to show any absorption, mostly because
they were $^{15}$NH lines. In support a lack of contamination we note
that in their analysis of the solar spectrum, Bi\'emont \etal{} (1982)
found that using the EW of the 3404\AA{} Pd line
resulted in a log$\epsilon$ of 1.67 dex, very close to the average
photospheric value of 1.69 dex and the meteoritic value of 1.70 dex
(Anders \& Grevesse 1989). In addition, several of our stars
(e.g. Fig 2) show very little absorption in that region, indicating
the lack of contaminants. Table 1 has our final line list. The
problem with blending is more severe for the Ag line at 3280\AA,
since Kurucz lists some NH lines very close to the wavelength of the
Ag line. We have set the NH strength in this region
empirically. Synthesis of the solar spectrum with the Kurucz line
lists revealed that the log gf-values for the 3280\AA{} region should
be increased by 0.4 dex relative to the log gf-values for the
3360\AA{} NH region. In practice, we found an NH abundance using the
3360\AA{} region for our halo stars, and then increased the N
abundance to 0.4 dex when synthesizing the 3280\AA{} region. The
linelist in Table 2 contains the original Kurucz values. We note that
if we did not adjust the NH absorption, the Ag values for HD 108577
and HD 186478 would be $\sim 0.1-0.15$ dex higher, still in
disagreement with the solar values. The stronger Ag line in BD +8 2548
makes it fairly immune to changes in the NH line strength. Ross \&
Aller (1972) and Crawford \etal{} (1998) found that the Ag line at
3280 in the Sun gives a log$\epsilon$ of 1.05 dex, below the
meteoritic value of 1.24 dex (Anders \& Grevesse 1989). The source of
the discrepancy between the meteoritic and the photospheric values is
unclear. Crawford \etal{} (1998) point out that an increase in the
opacity by $\sim$1.3 in the Sun could solve the problem. Whether
missing opacity on the same scale is present in our sample of
metal-poor giant is not known. We note that the gf-value is a recent
laboratory measurement (Fuhr \& Wiese 1996) and is estimated to have a
10\% uncertainty. Although HFS should not be a issue with a line this
weak, we have nonetheless used the wavelengths and relative strengths
from Ross \& Aller (1972) to split the Fuhr \& Wiese gf-value into HFS
components. We estimate that our signal-to-noise in these regions as
well as inaccuracies in our linelists limit our precision for both \ion{Pd}{1}
 and \ion{Ag}{1} to 0.2 dex. 
The inclusion of errors caused by our choices of model atmosphere
parameters leads to a final error estimate of 0.25 dex. Our Pd and Ag
abundances are listed in Table 2. In Table 2, we also list the
5-$\sigma$ upper limits on the abundance of Pd for the stars with
HIRES data. These limits were derived by generating synthetic spectra
with different Pd abundances and then testing the goodness-of-fit to
the HIRES data with a $\chi ^2$ test. The FWHM and the position of
the line were fixed; only the abundance was varied. The noise was
determined by the photon statistics. We also assumed that the
continuum was 99\% of our ``best'' choice and that Pd was the only
contributor to the absorption at that wavelength. These choices gave
a secure upper limit to the Pd abundance. Figure 2 gives an example
of the 5-$\sigma$ limit for BD $-$18 5550.

\begin{deluxetable}{lrrrrrrrrrrrrr}
\tablenum{3}
\tablewidth{0pt}
\tablecaption{Abundances}
\tablehead{
\colhead{Star} & \colhead{[Fe/H]} & \colhead{Sr}& \colhead{$\sigma$} 
& \colhead{Y}& \colhead{$\sigma$} 
& \colhead{Zr}& \colhead{$\sigma$} 
& \colhead{Pd}
& \colhead{$\sigma$} & \colhead{Ag} & \colhead{$\sigma$} 
& \colhead{Ba}& \colhead{$\sigma$} \\ 
\colhead{} & \colhead{} & \colhead{log$\epsilon$} & \colhead{} & \colhead{log$\epsilon$} &
\colhead{} & \colhead{log$\epsilon$} &\colhead{}
& \colhead{log$\epsilon$} &\colhead{}
& \colhead{log$\epsilon$} &\colhead{}
& \colhead{log$\epsilon$} &\colhead{}
}
\startdata
HD 29574 & $-$1.85 & 1.24 & 0.30 & 0.27 & 0.16 & 1.07 & 0.16 & \nodata & \nodata & \nodata & \nodata & 0.54 & 0.26 \\
HD 63791 & $-$1.72 & 1.07 & 0.30 & 0.28 & 0.17 & 1.01 & 0.15 & \nodata & 
\nodata & \nodata & \nodata & 0.42 & 0.26 \\
HD 88609 & $-$2.96 & $-$0.32 & 0.30 & $-$0.80 & 0.09 & 0.02 & 0.09 & 
$<$ \phs 0.07 & \nodata & \nodata & \nodata & $-$1.92 & 0.10 \\
HD 108577 & $-$2.36 & 0.35 & 0.30 & $-$0.52 & 0.13 & 0.18 & 0.10& 
$-$0.69 & 0.25 & $-$1.29 & 0.25 & $-$0.35 & 0.21 \\
HD 115444 & $-$3.14 & $-$0.51 & 0.30 & $-$1.00 & 0.09 & $-$0.30 &  0.08 & $<-$0.49 & \nodata & \nodata & \nodata & $-$1.10 & 0.15 \\
HD 122563 & $-$2.75 & $-$0.24 & 0.30 & $-$0.80 & 0.10 & $-$0.14  & 0.08  & 
$<-$0.60 & \nodata & \nodata & \nodata & $-$1.80 & 0.11  \\
HD 126587 & $-$3.07 & $-$0.28 & 0.30 & $-$1.07 &0.09 & $-$0.36 & 0.08 &
$<-$0.53 & \nodata & \nodata & \nodata & $-$1.08 & 0.16\\
HD 128279 & $-$2.38 & 0.06 & 0.30 & $-$0.72 & 0.15 & $-$0.09 & 0.14 & 
$<-$0.27 & \nodata & \nodata & \nodata & $-$0.74 &  0.17\\
HD 165195 & $-$2.31 & 0.68 & 0.30 & $-$0.38 &  0.08 & 0.46 & 0.11 & 
\nodata & \nodata & \nodata & \nodata & $-$0.43 & 0.20 \\
HD 186478 & $-$2.60 & 0.49 & 0.30 & $-$0.48 & 0.12 & 0.29 & 0.06 & 
$-$0.67 & 0.25 & $-$1.32 & 0.25 & $-$0.55 & 0.22 \\
HD 216143 & $-$2.22 & 0.87 & 0.30 & $-$0.16 & 0.14 & 0.53 & 0.10 &
\nodata & \nodata & \nodata & \nodata & $-$0.30 & 0.23 \\
HD 218857 & $-$2.18 & 0.61 & 0.30 & $-$0.43 & 0.13 & \nodata & \nodata
& \nodata & \nodata & \nodata & \nodata & $-$0.47 & 0.24 \\ 
BD $-$18 5550 & $-$3.04 & $-$1.20 & 0.30 & $-$1.81 & 0.05 & $-$1.22 & 0.09 & 
$<-$0.55 & \nodata & \nodata & \nodata & $-$1.67 & 0.16  \\
BD $-$17 6036 & $-$2.76 & $-$0.40 & 0.30 & $-$1.15 & 0.11 & $-$0.48 & 
0.09 & $<-$0.10 & \nodata & \nodata & \nodata & $-$1.09 & 0.18 \\
BD $-$11 145 & $-$2.48 & 0.31 & 0.30 & $-$0.57 & 0.11 & 0.09 & 0.14
& \nodata & \nodata & \nodata & \nodata & $-$0.29 & 0.23 \\ 
BD +4 2621 & $-$2.51 & 0.18 & 0.30 & $-$0.69 & 0.14 & 0.03 & 0.08 
& $<-$0.41 & \nodata & \nodata & \nodata & $-$1.21 & 0.23 \\
BD +5 3098 & $-$2.73 & 0.11 & 0.30 & $-$0.87 & 0.13 & $-$0.14 & 0.14 
& $<$ \phs 0.41 & \nodata & \nodata  & \nodata  & $-$0.96 & 0.18 \\
BD +8 2548 & $-$2.11 & 0.78 & 0.30 & $-$0.15 & 0.15 & 0.57 & 0.09 & 
$-$0.36 & 0.25 &  $-$0.78 & 0.25 & $-$0.07 & 0.24\\
BD +9 3223 & $-$2.28 & 0.71 & 0.30 & $-$0.23 & 0.14 & 0.44 & 0.17 
& \nodata & \nodata & \nodata & \nodata & $-$0.13 & 0.19 \\
BD +10 2495 & $-$2.07 & 0.77 & 0.30 & $-$0.16 & 0.14 & 0.48 & 0.17 
& \nodata & \nodata & \nodata & \nodata & 0.03 & 0.24 \\
BD +17 3248 & $-$2.10 & 0.94 & 0.30 & 0.10 & 0.16 & 0.75 & 0.18 
& \nodata & \nodata & \nodata & \nodata & 0.51 & 0.14 \\
BD +18 2890 & $-$1.73 & 1.21 & 0.30 & 0.38 & 0.16 & 1.03 & 0.14 
& \nodata & \nodata & \nodata & \nodata & 0.63 & 0.29 \\
\enddata
\end{deluxetable}

\subsection{Sr Abundances}

Another element useful for shedding light on early Galactic nucleosynthesis is
Sr. Because it is made in the r, main-s and weak s-processes, its
ratio with other elements, such as Y, changes as contributions from various
processes are made. Unfortunately, we found our synthesis of the strong resonance lines of SrII
at 4077\AA{} and 4215\AA{} yielded different answers for the
core and the wings of the lines.
The \ion{Sr}{2} line is very deep in most
of our stars, approaching depths of 20\% of the continuum. The high
layers of the atmospheric models probably have an incorrect
temperature structure. Indeed, the Fe lines argue that such is the case
(see Paper I) in these stars. So we decided to fit the wings of the
lines using the linelists from Sneden \etal{} (1996) to determine the 
Sr abundance. In the solar system, there
are four isotopes of Sr, $^{84}$Sr, $^{86}$Sr, $^{87}$Sr, and $^{88}$Sr, with
$^{88}$Sr dominating the abundance in the solar system and
$^{84}$Sr accounting for $<$ 1\%. Only $^{87}$Sr has
appreciable HFS. McWilliam \etal{} (1995) provide
the wavelengths and gf-values for the 4215\AA{} line; HFS constants
are not available for the 4077\AA{} line. The importance of HFS depends on
the relative strength of the $^{87}$Sr contribution. $^{88}$Sr is the
only Sr isotope produced in the r-process, so if the Sr in these stars
is due only to the r-process (as we argue below) then no HFS needs
to be considered. The main s-process produces mostly $^{88}$Sr, leaving another process,
most likely the weak s-process to contribute substantial amounts of
 $^{86}$Sr and $^{87}$Sr (Arlandini \etal{} 1999). To test 
the effect of including the weak s-process isotopes in our synthesis, we
subtracted Arlandini \etal 's main s-process yields from the total solar
system abundances, which resulted in a Sr composition that was 35\% $^{86}$Sr,
22\% $^{87}$Sr, and 43\% $^{88}$Sr. As suspected, using this combination
of Sr isotopes decreased the derived Sr abundance from the 
4215\AA{} line by up to 0.3 dex. Because this affected all of our
abundances in a similar fashion, the relative abundances change by 
a much smaller ($<$0.1 dex) amount. The wings of the lines are
not a sensitive abundance indicator; we estimate observational errors
of 0.15 dex for our abundances. Including the effect of uncertainties
in model atmosphere parameters, in particular in the microturbulent
velocity, raises the total error to 0.30 dex. Table 2 gives the Sr abundances
derived assuming only $^{88}$Sr is present.

\subsection{Other Abundances}

We have already presented the abundances for other neutron-capture
elements for the 22 stars in our sample in Paper I. 
Table 2 includes the [Fe/H] values derived in Paper I for
all our stars, as well as abundances for Y, Zr and Ba, since those
are the abundances most discussed in this paper. We have also plotted
some abundance ratios from the literature. When error bars for these points are
shown, they represent the addition in quadrature of the standard deviations
of the mean for the two elements.

\section{Results}

\subsection{Overview}

Figure 3 shows the abundances for three heavy-element-rich 
stars in our sample, as well as two extraordinary stars from the literature,
CS 22892-052 (Sneden \etal{} 1996) and CS 31082-001 (Hill \etal{} 2002).
The latter two are metal-poor ([Fe/H] $\sim -3.0$) field giants as
well, and show even larger enhancements of neutron-capture
elements than any of the stars in our sample. This permitted the measurement
of crucial elements such as U, Os, Ir, and Pb.
The elements heavier than Ba and lighter than Yb show remarkable star-to-star
consistency and very good
agreement with the contributions of the r-process to
the solar-system abundances (\rss).
The recent work of Toenjes \etal{} (2001) and Hill \etal{} (2002) 
on CS 31082-001 has shown that the
good agreement does not extend beyond Yb to Th and U. 
Our new Pd and Ag values
show a stronger odd-even effect than \rss, as seen by Sneden \etal{}
(2000a) in CS 22892-052. The additional abundances measured
in CS 22892-052 show that this odd-even pattern extends from Nb to Cd.
The lightest neutron-capture elements we have measured, Sr, Y, and Zr,
also show a similar pattern from star-to-star, although
even with the small number
of stars in Figure 3 the dispersion in the light-to-heavy
neutron-capture element ratios, such as Y/Ba, is obvious.
There is a lack of dispersion in the ratios between the intermediate elements, such as Pd and the heavy elements. 
\begin{figure}[b]
\begin{center}
\includegraphics[width=3in,angle=270]{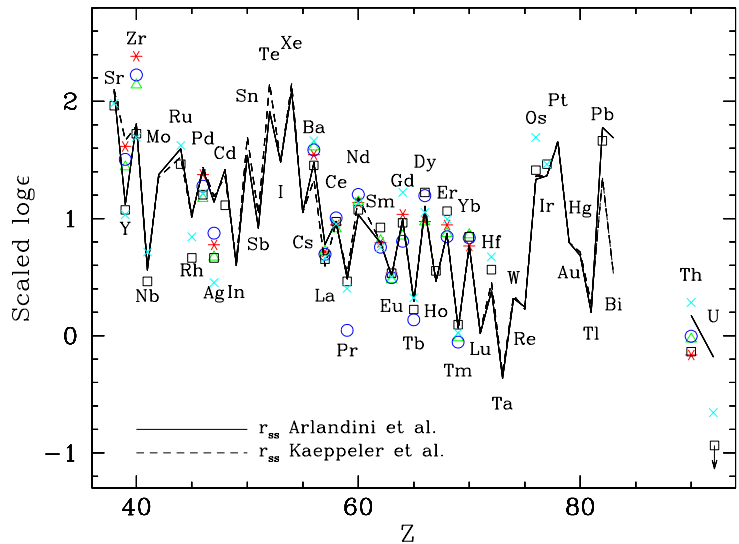}
\caption{\footnotesize Abundances from Sr to Yb for HD 108577 ($\triangle$),
HD 186478 ($\ast$) and BD +8 2548 ($\circ$) from this work. We have
also plotted the abundances for CS 22892-052 (Sneden \etal{} 2000a) (open 
squares) and CS 31082-001 (Hill \etal{} 2002) ($\times$). 
These abundances have been scaled up to the Arlandini \etal (1999) \rss{} 
using the Ba, La, Ce, Sm and Eu abundances. To show the uncertainties in
\rss, we have included the \rss{} from K\"appeler \etal{} (1989).}
\end{center}
\end{figure}

\begin{figure}[t]
\begin{center}
\includegraphics[width=3in,angle=0]{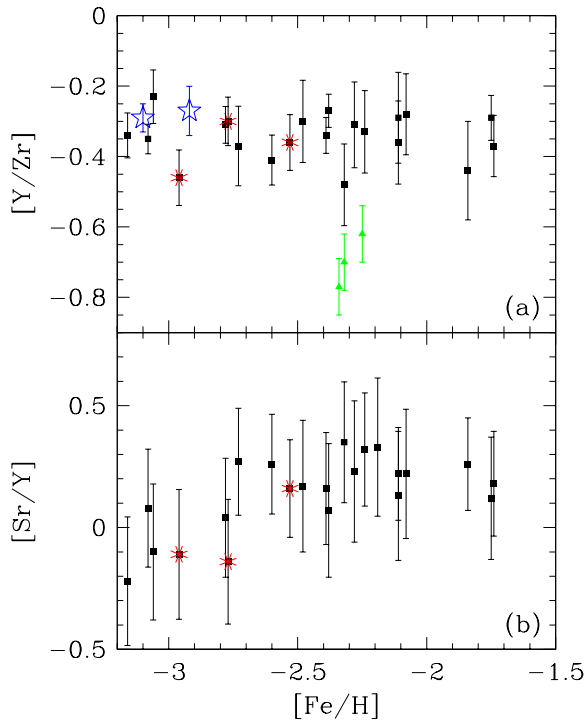}
\caption{\footnotesize (a) [Y/Zr] vs. [Fe/H] for stars in our sample (squares) 
as well as CS 22892-052 and CS 31082-001 (stars) and M 15 giants (triangles). 
Asterisks mark the three stars in our sample with
[Y/Ba]$>0.4$ dex. While there is good agreement among the [Y/Zr] values
for the field stars, the M 15 data are systematically lower. (b) [Sr/Y] vs.
[Fe/H] for stars in our sample. Since our Sr values were determined
using the wings of the lines, instead of the EW, we have not
included CS 22892-052 or CS 31082-001 in this plot. 
The lower values at lower metallicities
are due to difficulties of deriving abundances from the wings of the
lines. The observational errors are enough to explain all the scatter.} 
\end{center}
\end{figure}
\subsection{Ba through Yb}

From a quantitative standpoint, Johnson \& Bolte (2001) showed that 
the abundances of Ba through Yb in all stars 
agree well with \rss{} and that the scatter is consistent with observational
error. Previous
work on field stars with [Fe/H] $< - 2.0$
has given a similar result
(e.g. Gilroy \etal{} 1988; Sneden \etal{} 1996; McWilliam 1998). Sneden \etal{} (2000b) analyzed the abundances of
Ba, La, Ce, Nd, Sm, Eu, Gd and Dy in three red giants from the 
metal-poor globular cluster M 15, and found the \rss{} pattern in
each of these stars as well.
 
\subsection{Sr,Y, and Zr}

We also find that within the narrow mass range spanned by Sr, Y and Zr
(A=$84-96$), the abundance pattern repeats itself from star-to-star.
In Figure 4a, we have plotted [Y/Zr] vs. [Fe/H] and find that all the
scatter in the plot (aside from the M15 stars) is due to observational
error. A similar result for field dwarfs with [Fe/H] $< -1.5$ was
found by Zhao \& Magain (1991). We have included the three red giants
from M 15 (Sneden \etal{} 2000b) in Figure 4a. The M 15 stars have
similar [Y/Zr] values, but on average they are lower than the field
stars' values. The M 15 [Y/Zr] values are based on at least three
lines of each element, and the same source, but not exactly the same
lines, was used for the gf-values. So it seems unlikely that
observational error could provide the entire explanation. However,
observations of stars in more globular clusters, particularly at the
high S/N necessary to measure the \ion{Zr}{2} lines accurately, would
be helpful. The [Sr/Y] values shown in Figure 4b also are consistent
with one [Sr/Y], especially since the $\sim$0.3 dex difference between
the lowest metallicity stars and the rest of the sample is likely due
to the difficulties of determining abundances from the wings of lines
of varying strength. The plot looks substantially the same if we use
the abundances from the 4215\AA{} line with HFS and the isotope
percentages given in \S 3.2. \begin{figure}[h] \begin{center}
\includegraphics[width=3in,angle=0]{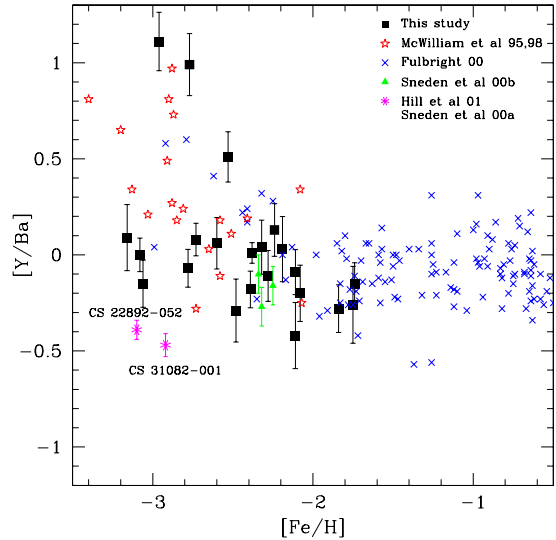} \caption{\footnotesize
[Y/Ba] for the stars from this study as well as from the literature.
There is a 1.5 dex scatter in [Y/Ba] which decreases as metallicity
increase. The two stars with extreme overabundances of the heavy
r-process elements, CS 22892-052 and CS 31082-001 mark the lower limit
of observed [Y/Ba] values. The M 15 stars, with their very small
dispersion in [Y/Ba], are marked by solid triangles.} \end{center}
\end{figure}

There is certainly scatter in the light-to-heavy
neutron-capture element ratios greater than can be explained by
observational error. Figure 5 shows the [Y/Ba] ratio for
our sample as well as for data from McWilliam \etal{} (1995), McWilliam
(1998) and Fulbright (2000) taken from the literature. The literature
studies were drawn from recent papers that had data with high-resolution 
and large
wavelength coverage of metal-poor stars.
Our results (Figure 5) confirm the large scatter seen previously in the ratio
of light-to-heavy neutron capture elements (e.g. McWilliam \etal{} 1995; 
Ryan \etal{} 1996; Westin \etal{} 2000)
While we have chosen to plot [Y/Ba] because their lines are prominent in
all of our spectra, we note that using Sr or Zr in
place of Y, or any element from La to Yb in place of Ba would have
resulted in a similar plot, since the abundance ratios within each group
are constant.

\subsection{Pd and Ag}
Does a similar scatter exist in the [Pd/Y] or [Pd/Ba] values? 
For the three stars in our sample with measurements, the 
[Pd/Y] and [Pd/Ba] values are consistent with a single value. 
The 5-$\sigma$ upper limits quoted for Pd cannot rule out that all
stars have the ratio for these elements. Higher S/N data, especially of
stars with high [Y/Ba] ratios such as HD 88609 and HD 122563, would
answer this question. 
We find a similar [Pd/Ag]
ratio to the values for CS22892-052 (Sneden \etal{} 2000a) and
CS 31082-001 (Hill \etal{} 2002) in the three stars in our
sample with measurements. It is important to keep in mind that this
sample is biased to stars with large [heavy-neutron-capture/Fe] ratios and
may not be representative of all stars. However, the consistent [Pd/Ag]
ratio we are finding may indicate one r-process pattern in the
intermediate mass elements in metal-poor stars.

\subsection{Summary}
Our observational results can be summarized as follows:  There is
good agreement between Ba and Yb with \rss{} for all the stars, both
field and cluster, in our sample.
For the small number of stars with Pd and Ag measurements, we find
indications of an increased odd-even effect for Pd and Ag compared to 
\rss. There is a large spread in [Y/Ba], but all the field stars have
similar values for [Y/Zr] and [Sr/Y]. However, the [Y/Zr] values
for stars in M 15 are $\sim$ 0.3 dex lower.

\section{Discussion}
\subsection{Origin of the neutron-capture elements in metal-poor stars}
\subsubsection{Uncertainities in theoretical r-process abundances}
Ideally, the abundance ratios produced by the r-process
would be known accurately from theory, 
and the results from metal-poor stars could
be interpreted in light of that information.
But because of the unknown physical properties
of the progenitor nuclei near the neutron-drip line and the unknown
physical conditions during the r-process, the most useful predictions
actually rely on s-process calculations. These s-process models are fit
to the s-only nuclei and then subtracted from the total isotopic
abundances present in the solar system. 
Therefore, depending on the results of the s-process model, \rss{} can change. This change can be substantial
for elements that in the solar system are mostly produced by the
s-process, such as Sr, Y, and Zr, where changes of 10\% in the
s-process predictions can lead to changes of over 100\% in
\rss{}. 
We have illustrated the consequences in Figure 3. Here
we have plotted \rss$={\rm log} \epsilon_{tot}- 
{\rm log} \epsilon_{main s-process}$ from two literature sources: K\"appeler
\etal{} (1989) and Arlandini \etal{} (1999). 
The main difference between the two studies lies in the
physical conditions during the neutron-captures onto the seed nuclei.
K\"appeler \etal{} used the ``classical'' model, where the neutron 
exposure is assumed to have an exponential distribution and the temperature
does not depend on time.
 Arlandini \etal{} instead use
the physical conditions predicted by AGB models, where neither assumption
holds.
This substantially
modifies the abundances, particularly of Y. Figure 3 shows
that if the Arlandini results, instead of the K\"appeler results,
are used, there is no conflict between
\rss{} and the abundances of CS 22892-052 in the Sr, Y and Zr region
This does not mean that the problem of the abundance ratios of the
lighter neutron-capture elements has been solved, but rather that
uncertainties in \rss{} remain at a large enough level to confuse
our interpretations of the abundances in metal-poor stars. 
\subsubsection{Theoretical predictions for the r-process}
Theoretical results show that the r-process alone is responsible for
the production of the elements heavier than Zr in the early Galaxy. 
The main s-process is not efficient at low metallicities (Gallino 
\etal{} 1998). Combined with the lifetimes of the progenitor low-mass AGB stars, this leads to
an 0.7 Gyr time lag to the appearance of s-process nucleosynthesis
in the early Galaxy (Raiteri \etal{} 1999).
No
other process contributes significant amounts of the heavier nuclei.
The main s-process can never build the heaviest nuclei, such as Th and
U (Clayton \& Rassbach 1967) so these must be r-only nuclei at all
times. 
Kratz \etal{} (2000) found that requiring a high minimum neutron density during
the r-process could duplicate the
 enhanced odd-even staggering in the Pd-Ag region and still match
the pattern seen in the heavier elements. Sr, Y and Zr
are produced in the r-process and are also
seeds for the creation of the heavier elements such as Pd, Ba, Yb, etc. 
In fact, if the neutron density is high enough,  
it is possible for so many neutrons to be captured that
little Sr, Y, and Zr is produced. 
(Kratz \etal{} 2000;  Pfeiffer, Ott \& Kratz 2000). Depending
on the physical conditions in the r-process event, very different
[Y/Ba] ratios, for example, can be created. Thus, it is possible
for the r-process to explain all the features we observe.

\subsubsection{Theoretical predictions for the weak s-process}
However, the r-process may not be the only contributor to the light
neutron-capture elements. 
The weak s-process, which
occurs in short-lived, massive stars, is another possible source for 
Sr, Y, Zr in low-metallicity stars. As mentioned earlier, the conditions
in massive stars do not lead to the production of heavier nuclei
than A$\sim$90, so contributions from the weak s-process are not
expected for elements heavier than Zr. Prantzos \etal{} (1990) argued that 
the weak s-process
should be extremely inefficient at [Fe/H]$\sim -3.0$ because of 
the lack of seed nuclei
and the presence of primary neutron poisons. However,
there are several uncertain quantities in those calculations, particularly
crucial neutron capture cross-sections, which could alter the
efficiency of the weak s-process in metal-poor stars by factors of $\sim$ 5
(Prantzos \etal{} 1990; Rayet \& Hashimoto 2000), so contributions
to Sr, Y, and Zr from the weak s-process cannot be dismissed.

\subsubsection{Constraints from observations: the r-process}
The observed good agreement between the heavier elements (Ba-Yb) and
\rss{} is empirical evidence for the predominance of the r-process in
the early Galaxy. Adding contributions from the main s-process to
\rss{} results in a poor fit with the abundances in metal-poor stars
(Gilroy \etal{} 1988; Sneden \etal{} 1996; Johnson \& Bolte 2001)
Goriely \& Arnould (1997) showed that the pattern from Ba to Yb
is not very sensitive to the exact conditions of temperature, entropy, etc.
in which the r-process takes place, so this abundance pattern is perhaps
a robust sign of the r-process. 

Observations also show that the r-process contributes at least some
fraction to the lighter neutron-capture elements.
First, we note that the values for [Y/Zr] and [Sr/Y] that
we find are close to the those we would predict if we assumed
that all possible r-process isotopes were produced in equal amounts
as discussed by Goriely \& Arnould. Y has one stable
r-process isotope, Zr has five, and Sr has one. This leads to 
[Y/Zr]$=-0.34$ and [Sr/Y]$=0$, similar to what is seen in Figure 4.

Then there are the cases of CS 22892-052 and CS 31082-001. 
These two stars show large enhancements of the heavy neutron-capture
elements, such as Ba, which are due to the r-process (Sneden \etal{} 1996).
They also are enriched in Sr, Y and Zr to a similar degree.  
It is more likely that both the Ba and
the Y were created in the same r-process event than 
than that these two stars were enriched by an r-process event
that created Ba, but not Y, and were also enriched in Y by a separate, 
weak s-process event.
(McWilliam 1998).
Even better evidence comes from M15. 
Like most other Galactic globular clusters,
stars in M 15 are homogeneous in their chemical composition when considering
the $\alpha -$ and iron-peak elements, with the exception of elements that
have been affected by mixing (Sneden et al. 1997). However, the
[Ba/Fe] values showed a spread of 0.8 dex. Examination of other
heavy neutron-capture elements, such as Eu,
showed that the heavy neutron-capture elements in all the stars was
due to the r-process (Sneden \etal{} 2000b).
Somehow, the r-process ejecta had managed to spread itself very unevenly in
this cluster.
In Figure 5, we have included
the [Y/Ba] from the three stars of Sneden et al. (2000b). The values for these 
stars are indistinguishable given the observational errors. It is
unreasonable to expect an independent process to have dispersed its
ejecta in exactly the same pattern as the r-process event. So the Y in
CS 22892-052, CS 31082-001 and M15 comes from the r-process.

\subsubsection{Constraints from observations:  the weak s-process}
We can then use the knowledge that the Y in these stars 
was produced in the
r-process in the early Galaxy, to evaluate the possible contributions
of other processes to the lighter neutron capture elements. 
The [Y/Zr] ratio is sensitive to contributions from different
processes. In
Figure 4,  we saw that stars 
have the same [Y/Zr] regardless of their [Y/Ba] value; the three stars
from our sample that had high [Y/Ba] values in Figure 5 are hidden
amidst the rest of the sample. Because the [Y/Zr] ratio is
very sensitive to the production method, as we discuss in detail
below,   
the constant [Y/Zr] values make 
it very difficult to attribute the high [Y/Ba] values observed to the
addition of light elements from another process that was not the
r-process.

Let us
take the case of CS 22892-052 ([Y/Ba]=$-0.39$) and HD 122563 ([Y/Ba]=0.99).
If we first assume that CS 22892-052's [Y/Ba] represents the 
minimum production in an r-process event, we then find that 96\%
of the Y in HD 122563 must have been produced in another process. 
The weak s-process yields of Raiteri \etal{} (1993) predict a
[Y/Zr] value for HD 122563 of 0.25, some 0.55 dex larger 
than measured. The situation is not significantly altered by having some
of the Y in both stars from the weak s-process.  It is very difficult
to hide even a small contribution from the weak s-process when
[Y/Zr] is constant. This is because 
with a predicted production [Y/Zr] production ratio of 0.43 dex, the
weak s-process is very different from the [Y/Zr] expected from
the r-process ($-$0.33 according to Arlandini \etal (1999)). 
It is unclear whether the predictions of Raiteri \etal{} are
valid at all points in Galactic history.
In order to understand our observational results, it is crucial to
know the expected range of [Y/Zr] values for other processes that
could contribute to the the heavy elements in the early Galaxy. 
If there are other processes, their [Y/Zr] production ratio
must be similar to that of the r-process. The simplest solution, however,
is to have the
production of all neutron-capture elements in the r-process only 
in the early
Galaxy. 

\subsection{The r-process in the early Galaxy}

Although at present we favor a picture where 
all the neutron-capture elements in metal-poor stars
were created in the r-process, the spread in [Y/Ba] shows
that there is not one universal r-process pattern. 
The abundances in CS22892-052 and CS31082-001 have already shown
that (Sneden \etal{} 2000a; Hill \etal{} 2002). 
In addition, Wasserburg \etal{} 
(1996) found evidence in the abundances of extinct radioactive nucleotides
in the solar system that some r-process events produce much larger
$^{129}{\rm I}/^{182}{\rm Hf}$ ratios than others. 
Because of the scatter in [Y/Ba], the abundances in metal-poor stars give
support for some r-process events producing mostly the lighter neutron-capture
elements, while other events favor the production of Ba and heavier.
Some events also appear to manufacture more of the heaviest elements
such as Th and U. Such an event must have polluted CS 31082-001.
Whether one phenomenon, such as the neutrino-wind in 
Type II SN, can provide a wide enough range of conditions to account
for the diversity seen, or whether a variety of phenomena, from
neutron-star mergers to He-burning regions, are needed is not yet clear.

\section{Conclusion}
Abundance ratios 
of neutron-capture elements in metal-poor stars show both striking
similarities and large dispersions. There is real scatter 
in the [Y/Ba] ratios observed and differences 
between metal-poor stars and \rss{} in the [Pd/Ag] ratio. 
The Sr-Y-Zr and the Ba$-$Yb regions
show similar abundance patterns in all stars in the Paper I sample, with
some deviation in the lighter elements in the red giants from M 15.
The M 15 giants provide the strongest evidence that the light neutron-capture
elements in metal-poor stars
can be produced in the same events as the heavy neutron-capture elements, 
since stars with very different
[Ba/Fe] have identical [Y/Ba]. Similar evidence is provided by
CS 22892-052 and CS 31082-001, which have large enhancements in both
Y and Ba.

Theoretical results show that the main s-process cannot produce substantial
amounts of the neutron-capture elements in low metallicity stars. This
result is supported by the good agreement between 
abundance pattern in the Ba region in our
sample of stars and \rss. The weak s-process potentially could contribute
to elements with A$<90$. However,  
the constant abundance ratios of [Sr/Y] and [Y/Zr] precludes
substantial contributions from processes other than the r-process in
the early Galaxy. 
As a result, the abundances of the neutron-capture elements in metal-poor stars provide
strong constraints on the r-process. Any model of the r-process must
explain the scatter seen in [Y/Ba] and [Ba/Th]. In addition,
the models need to reproduce the enhanced odd-even effect in the Pd-Ag region.
The variety of phenomena proposed for the r-process shows that the r-process
production need not be confined to one kind of event, which could
aid in describing the dispersion seen.

\acknowledgements{Some of the data presented herein were obtained at the W.M. Keck Observatory, which is operated as a scientific partnership among the California Institute
of Technology, the University of California and the National Aeronautics and Space Administration. The Observatory was made possible by the generous
financial support of the W.M. Keck Foundation.  
We would like to thank Andy McWilliam for a careful
reading of a draft of this paper and Chris Sneden for generously making
software and linelists available. This work was supported by
NSF AST-0098617.}


\clearpage
\begin{references}
\vspace{.1in}
\reference{} Anders, E. \& Grevesse, N. 1989, Geo. Cos. Acta, 53, 503
rawford, J. 

\reference{} Arlandini, C. K\"appeler, F., Wisshak, K., Gallino, R., 
Lugaro, M., Busso, M., \& Straniero, O. 1999, ApJ, 525, 886

\reference{} Biemont, E., Grevesse, N., Kwiatkowski, M., \& Zimmermann, P.
1982, A\&A, 108, 127

\reference{} Burbidge, E. M., Burbidge, G. R., Fowler, W. A., \& Hoyle, F.
1957, Rev. Mod. Phy., 29, 547

\reference{} Burris, D. L., Pilachowshi, C. A., Armandroff, T. E., Sneden, C.,
Cowan, J. J., \& Roe, H. 2000, ApJ, 544, 302

\reference{} Cameron, A. G. W. 1957, PASP, 69, 201

\reference{} Clayton, D. D. \& Rassbach, M. E. 1967, ApJ, 148, 69

\reference{} Couch, R. G., Schmiedekamp, A. B., \& Arnett, W. D. 1974, ApJ, 190, 95

\reference{} Cowan, J. J., Cameron, A. G. W., \& Truran, J. W. 1985, ApJ, 294, 656

\reference{} Crawford, J. L., Sneden, C., King, J. R., Boesgaard, A. M.,
\& Deliyannis, C. P. 1998, AJ, 116, 2489

\reference{} Freiburghaus, C., Rembges, J.-F., Rauscher, T., Kolbe, E.,
Thielemann, F.-K., Kratz, K.-L., Pfeiffer, B., \& Cowan, J. J. 1999, ApJ,
516, 381

\reference{} Freiburghaus, C., Rosswog, S., \& Thielemann, F.-K. 1999, ApJ, 525, L121

\reference{} Fuhr, J. R. \& Wiese, W. L. 1996, in CRC Handbook of Chemistry
and Physics, ed. D. R. Lide (Boca Raton, FL: CRC Press), 10-128

\reference{} Fulbright, J. 2000, AJ, 120, 1841

\reference{} Gallino, R., Arlandini, C., Busso, M., Lugaro, M.,
Travaglio, C., Straniero, O., Chieffi, A., \& Limongi, M. 1998, ApJ, 497,
388

\reference{} Gilroy, K. K., Sneden, C. Pilachowski, C., \& Cowan, J. J. 1988,
ApJ, 327, 298

\reference{} Goriely, S. \& Arnould, M. 1997, A\&A, 322, L29

\reference{} Hill, V., et al. 2002, A\&A, 387, 560

\reference{} Johnson, J. A., 2002, ApJS, 139, 219 

\reference{} Johnson, J. A. \& Bolte, M. 2001, ApJ, 554, 888

\reference{} K\"appeler, F., Beer, H, \& Wisshak, K. 1989, Rep. Prog. Physics,
52, 945


\reference{} Kratz, K.-L., Pfeiffer, B., Thielemann, F.-K., \& Walters, W. B.
2000, Hyperfine Interactions, 129, 185 (astro-ph/9907071)

\reference{} Kurucz, R. L. \& Bell, B. 1995, 1995 Atomic Line Data, CD-ROM 23,
Cambridge, MA: Smithsonian Astrophysical Observatory.

\reference{} Lattimer, J. M. \& Schramm, D. N. 1974, ApJ, 192, L145

\reference{} McWilliam, A. 1998, AJ, 115, 1640

\reference{}McWilliam, A., Preston, G. W., Sneden, C. \& Searle, L. 1995, 
AJ, 109, 2757

\reference{} O'Brian, T. R., Wickliffe, M. E., Lawler, J. E., Whaling, W., \& Brault, J. W> 1991, J. Opt. Soc. Am. B, 8, 1185

\reference{} Otsuki, K., Tagoshi, H., Kajino, T., \& Wanajo, S. 2000, ApJ, 533, 424

\reference{} Pfeiffer, B., Ott, U., \& Kratz, K.-L. 2001, Nuc. Phy. A. 688, 465

\reference{} Pickering, J. C. 1996, ApJS, 107, 811

\reference{} Prantzos, N., Hashimoto, M., \& Nomoto, K. 1990, A\&A, 234, 211

\reference{} Qian, Y.-Z. \& Woosley, S. E. 1996, ApJ, 471, 331

\reference{} Raiteri, C. M., Gallino, R., Busso, M., Neuberger, D., \& K\"appeler, F. 1993, ApJ, 419, 207

\reference{} Raiteri, C. M., Villata, M., Gallino, R., Busso, M., \& Cravanzola, A. 1999, ApJ, 518, L91

\reference{} Rayet, M. \& Hashimoto, M. 2000, A\&A, 354, 740

\reference{} Ross, J. E. \& Aller, L. H. 1972, Sol. Phy., 25, 30

\reference{} Rosswog, S., Davies, M. B., Thielemann, F.-K., \& Piran, T. 2000,
A\&A, 360, 171

\reference{} Ryan, S. G., Norris, J. E., \& Beers, T. C. 1996, ApJ, 471, 254

\reference{} Sneden, C., Cowan, J. J., Ivans, I. I., Fuller, G. M., Burles, C.,
Beers, T. C., \& Lawler, J. E. 2000a, ApJ, 533, L139

\reference{} Sneden, C., Kraft, R. P., Shetrone, M. D., Smith, G. H.,
Langer, G. E., \& Prosser, C. F. 1997, AJ, 114, 1964

\reference{} Sneden, C., Johnson, J., Kraft, R. P., Smith, G. H., Cowan, J. J., \& Bolte, M. S. 2000b, ApJ, 536, L85

\reference{} Sneden, C., McWilliam, A., Preston, G. W., Cowan, J. J., Burris,
D. L., \& Armosky, B. J. 1996, ApJ, 467, 819

\reference{} Takahashi, K., Witti, J., \& Janka, H.-Th. 1994, A\&A, 286, 857

\reference{} Toenjes, R., Schatz, H., Kratz, K.-L., Pfeiffer, B., Beers, T. C., Cowan, J. \& Hill, V. 2001, in Astrophysical Ages and Timescales, ASP 
Conf. Series Vol 245, Ed T. von Hippel, C. Simpson and N. Manset
San Francisco, ASP, p. 376 astroph/010433

\reference{} Truran, J. W., 1981, A\&A, 97, 391

\reference{} Truran, J. W., Cowan, J. J., \& Fields, B. D. 2001, Nuc. Phy. A, 
688, 330

\reference{} Vogt, S. S., 1987, PASP, 99, 1214

\reference{} Vogt, S. S., et al., 1994, SPIE, 2198, 362

\reference{} Wanajo, S., Kajino, T., Mathews, G. J., \& Otsuki, K. 2001, ApJ,
554, 578

\reference{} Wasserburg, G. J., Busso, M., \& Gallino, R. 1996, ApJ, 466, L109

\reference{} Westin, J., Sneden, C., Gustafsson, B., \& Cowan, J. J. ApJ, 530, 783

\reference{} Witti, J., Janka, H.-Th., \& Takahashi, K. 1994, A\&A, 286, 841

\reference{} Woosley, S. E. \& Hoffman, R. D. 1992, ApJ, 395, 202

\reference{} Woosley, S. E., Wilson, J. R., Mathews, G. J., Hoffman, R. D., \& Meyer, B. S. 1994, ApJ, 433, 229

\reference{} Zhao, G. \& Magain, P. 1991, A\&A, 244, 425

\end{references}
\end{document}